\documentstyle[prl,aps,epsf,twocolumn]{revtex}

\title
{
A Closed Class of Hydrodynamical Solutions for the Collective
Excitations of a Bose-Einstein Condensate
}

\author
{
Pippa Storey\cite{permanentaddress}$^1$  and Maxim Olshanii$^{1,2}$
}
\address
{
$^1$Lyman Laboratory, Harvard University,
Cambridge, MA 02138, USA\\
$^2$Department of Physics and Astronomy,
University of Southern California,
Los Angeles CA 90089-0484 USA\\
{\small E-mail: {\it p.storey@auckland.ac.nz, maxim@atomsun.harvard.edu}}
}

\date{\today}

\begin{document}
\maketitle
\begin{abstract}
A trajectory approach is taken to the hydrodynamical treatment of
 collective excitations of  a Bose-Einstein condensate in a 
harmonic trap. The excitations induced by linear
deformations of the trap are shown to constitute a broad class of solutions
that can be fully described by a simple nonlinear matrix equation. 
An exact closed-form expression is obtained for the
solution describing the mode  $\left\{ n=0,m=+2 \right\}$
in a cylindrically  symmetric trap, and 
the calculated amplitude-dependent frequency shift shows good
agreement with the experimental
results of the JILA group.
\end{abstract}
%
%




The recent realisation of Bose-Einstein condensation in dilute trapped
atomic vapours \cite{Anderson,Bradley,Davis} has motivated an extensive
theoretical study of these systems (for a review see reference
 \cite{Dalfovo}). Many of the nonlinear features of the condensates
are manifested in their collective excitations
\cite{Cornell_1,Cornell_2,MIT_1,MIT_2}. In this letter we
consider the collective excitations of a condensed atomic gas 
in a harmonic trap in the 
regime in which  the number of atoms is sufficiently
large that the  hydrodynamical treatment of
Stringari  is valid \cite{Stringari}. We apply this treatment by
means of a trajectory approach, which constitutes a generalisation
of the scaling transformation of Castin and Dum \cite{CastinDum}.
To first order in the excitation amplitude our results for
the oscillatory modes in a stationary trap are identical to those 
of the perturbative treatment of
\"Ohberg {\em et al.} \cite{Shlyapnikov}. However for a broad class
of excitations, our analysis produces a simple
nonlinear matrix equation, which is valid to all orders in the excitation
amplitude. The excitations to which this matrix equation applies
are those induced by time-varying
linear deformations of the trap (that is, perturbations
of the trapping potential that maintain its harmonicity). 
This class includes all the  excitations that
have to date been studied experimentally. 

We suppose that the condensed gas is confined by
a harmonic potential, which, in its most general form, is given by
\begin{equation}
U({\bf r},t)=\frac{1}{2}\sum_{i,j} 
{\cal K}_{ij} (t)
 \left[r_i- \bar{r}_i (t)\right] \left[r_j- \bar{r}_j (t)\right]
  ,\label{e_genpot}
\end{equation}
where $\bar{\bf r}(t)$ is the position of the centre of the trap,
and ${\cal K}(t)$ is a symmetric matrix with components
${\cal K}_{ij}(t)$, which represent the spring `constants' of the trap.

In the mean field approximation, the evolution of the condensate wavefunction
is determined by the time-dependent Gross-Pitaevski equation, which, for
zero temperature, is
\begin{equation}
i \hbar \partial_t \Phi\left({\bf r},t\right)= 
\left[-\frac{\hbar^2}{2m}\triangle
+  U({\bf r},t) + N g \left| \Phi\left({\bf r},t\right) \right|^2 \right]
\Phi\left({\bf r},t\right)  \label{e_GP}
\end{equation}
where $N$ is the number of atoms in the condensate, and $g$ is a measure of
the strength of the atomic interactions.
 $g$ is related to the  $s$-wave scattering length $a$  through
$g = 4 \pi \hbar^2 a / m$, and is assumed to be positive.

Defining the density of the condensate as $\rho \left({\bf r},t\right) = N
  \left|\Phi \left({\bf r}, t\right) \right|^2$, and its velocity field
as 
$
{\bf v}\left({\bf r},t\right)= (\hbar/2im) \, 
\mbox{\boldmath $\nabla$} 
\log\lbrack\Phi({\bf r}, t) / \Phi^{\ast} ({\bf r}, t) \rbrack
$,
it is easily shown that the
 Gross-Pitaevski equation (\ref{e_GP}) is equivalent  to the following equations
for $\rho$ and ${\bf v}$
\begin{eqnarray}
\partial_t \rho + \mbox{\boldmath $\nabla$}
 \left( {\bf v} \rho \right) & = & 0 \\
m \partial_t  {\bf v} + 
\mbox{\boldmath $\nabla$} \left( U_{\text{eff}} 
+ 
\frac{1}{2} m v^2 \right) & = & 0 ,
\end{eqnarray}
where
\begin{equation}
U_{\text{eff}}=  U + g \rho -
\frac{\hbar^2}{2 m \sqrt{\rho}} \triangle \sqrt{\rho} 
\end{equation}
is an effective potential.

Following Stringari \cite{Stringari} we consider the limit in
which the number of atoms $N$ is sufficiently large that the
density $\rho$ becomes smooth, and it is valid to neglect the
kinetic energy pressure term 
$\hbar^2 \left(\triangle \sqrt{\rho}   \right)/2 m \sqrt{\rho}$.
The ratio between the typical force caused by this force and 
the force $-\mbox{\boldmath $\nabla$}(g\rho)$ caused by inter-particle 
interactions can be shown to be of the order 
of $(\hbar^2/mR_{0}^2)/\mu \sim N^{-2/3} (T_{c}/\mu)^2$,
and for the most of the BEC experiments this ratio is 
as small as $10^{-1}\!-\!10^{-3}$. (Here and below 
$R_{0} \sim \sqrt{\mu/m\omega^2}$ is the Thomas-Fermi 
radius of the condensate, $T_{c} \sim \hbar\omega N^{1/3}$ is the 
Bose-Einstein condensation temperature, $\mu \sim g\rho$ is the  
chemical potential, and $\omega$ is a typical trapping frequency.) 
In this regime the condensate can be
modeled as a classical gas in which each particle
is subject to a force
\begin{equation}
{\bf F} \left({\bf r},t\right)=- 
\mbox{\boldmath $\nabla$}_{\bf r} \left[  U({\bf r},t)
+ g \rho \left({\bf r},t\right)  \right] , \label{e_force}
\end{equation}
and for which the velocity field
is irrotational. We include the subscript ${\bf r}$ on the
gradient operator $\mbox{\boldmath $\nabla$}$
to indicate explicitly that all derivatives are
taken with respect to the components of ${\bf r}$.
This notation is used since we shall later introduce
a change of coordinates.

The requirement that
 $\mbox{\boldmath $\nabla$}_{\bf r} \times {\bf v}(t) \equiv 0$
 imposes a constraint only on the initial condition,
the reason being that 
since  the  only force involved  (\ref{e_force})
is the gradient of a potential, the total derivative of
$\mbox{\boldmath $\nabla$}_{\bf r} \times{\bf v}(t) $ vanishes.
This means that, provided the
velocity field is initially irrotational everywhere, it will remain
so with time.

Using expression    (\ref{e_force})
for the classical force, we obtain the following evolution equation
for ${\bf r}(t)$
\begin{equation}
m \ddot{\bf r}(t)= - {\cal K}(t)\left[ {\bf r}(t)-\bar{\bf r}(t) \right]
- g \mbox{\boldmath $\nabla$}_{\bf r}
\rho \left({\bf r},t\right)\label{e_ddotr} .
\end{equation}
In order to solve this equation, it is necessary to
determine the density of the condensate at position
${\bf r}$ and time $t$, which in turn requires a knowledge
of the trajectories of all the other particles in the condensate.
To this end, we associate with each trajectory 
${\bf r}(t)$ a unique reference point
 ${\bf r}_0$,  corresponding to the position of a particle
in the ground state  $\Phi_0$ of the condensate 
in the hydrodynamical limit
 for a stationary harmonic  trap $U_0 ({\bf r})$, given by equation
(\ref{e_genpot}) with  $\bar{\bf r}(t)=0$ and ${\cal K}(t)={\cal K}_0$.
 $\Phi_0$ is obtained from the solution 
 $\rho_0 $  of equation (\ref{e_force}) with
${\bf F} \left({\bf r},t\right) = 0$, through
\begin{equation}
\rho_0 \left({\bf r}_0 \right) = N \left|\Phi_0 \left({\bf r}_0 \right)
\right|^2 , \label{e_TF}
\end{equation}
and is given by
\begin{equation}
\Phi_0 \left({\bf r}_0 \right)= \sqrt{\frac{\mu- U_0({\bf r}_0)}{Ng}} ,
\end{equation}
where $\mu$ is the chemical potential. Note that this
is just the Thomas-Fermi solution.

The density of the condensate
at time $t$ can now  be written as
\begin{equation}
\rho \left({\bf r},t\right) = 
\det \left[\frac{\partial {\bf r}}{\partial {\bf r}_0}
 \right]^{-1}
\rho_0 \left({\bf r}_0 \right) . \label{e_rhot}
\end{equation}
where $\partial {\bf r} /\partial {\bf r}_0 $ is a matrix
whose $ij$ th element is  $\partial  r_i /\partial  r_{0,j} $.
 Given that $ \mbox{\boldmath $\nabla$}_{\bf r} = 
\left[ \left({\partial {\bf r}}/
{\partial {\bf r}_0}\right)^{-1}\right]^T 
\mbox{\boldmath $\nabla$}_{{\bf r}_0} $,
we can rewrite equation (\ref{e_ddotr}) as a partial
 differential equation for 
 ${\bf r}({\bf r}_0,t) $
\begin{eqnarray}
m \partial^2_t {\bf r} & = &- {\cal K}(t)
\left[ {\bf r}-\bar{\bf r}(t)\right] \nonumber \\
& & 
- g \left[ \left(\frac{\partial {\bf r}}
{\partial {\bf r}_0}\right)^{-1} \right]^T \mbox{\boldmath $\nabla$}_{{\bf r}_0}
\left[\det \left(\frac{\partial {\bf r}}
{\partial {\bf r}_0} \right)^{-1}
\rho_0 \left({\bf r}_0 \right) \right] . \nonumber \\
&&  \label{e_genevol}
\end{eqnarray}
The functions ${\bf r}$ satisfying this equation can be used to
construct hydrodynamical solutions of the 
Gross-Pitaevski equation (\ref{e_GP}). Given that
the velocity field is irrotational, we can
express it as the gradient of some 
function $\Theta\left({\bf r},t \right)$
\begin{equation}
{\bf v}(t) = \mbox{\boldmath $\nabla$}_{\bf r} \Theta\left({\bf r},t \right) ,
\end{equation}
in terms of which we obtain
\begin{equation}
 \Phi\left({\bf r},t\right) \approx e^{-i \beta(t)} 
e^{im \Theta \left({\bf r},t \right)/\hbar} 
\frac{\Phi_0\left({\bf r}_0\right)}
{\sqrt{\det \left( \partial {\bf r}/
\partial {\bf r}_0  \right)}}, \label{e_gensoln}
\end{equation}
where ${\bf r}_0$ is uniquely defined in terms of 
 ${\bf r}$ and $t$, and $\beta(t)$ is some function of time.

By applying this analysis to oscillatory modes in the stationary 
harmonic potential $U({\bf r},t)=U_0({\bf r})$
(that is, ${\cal K}(t) ={\cal K}_0 $, $\bar{\bf r}(t)=0$)
in the limit of small excitation amplitudes, we recover the results of 
  \"Ohberg {\em et al.} \cite{Shlyapnikov}, as we now show.
For low amplitude oscillations 
\begin{equation}
{\bf r}\left({\bf r}_0,t \right) \approx {\bf r}_0 +
\epsilon \left[ \delta{\bf r}\left({\bf r}_0 \right) e^{-i \omega t}+
\delta {\bf r}^{\ast}\left({\bf r}_0 \right) e^{i \omega t}\right],  \label{e_calP}
\end{equation}
where $\epsilon$ is a perturbation parameter. Substituting this expression
into equation (\ref{e_genevol}), and keeping only first order terms  in $\epsilon$
we obtain the relation
\begin{equation}
m \omega^2\, \delta {\bf r} = \mbox{\boldmath $\nabla$} \left\{ \delta{\bf r} \cdot 
\left({\cal K}_0 {\bf r} \right)-
\left[ \mu- U_0({\bf r}) \right] 
\mbox{\boldmath $\nabla$} \cdot \delta {\bf r}
 \right\} ,\label{e_P}
\end{equation}
in which we have dropped the distinction between ${\bf r}$ and
${\bf r}_0$, since this affects only higher orders.
$\delta {\bf r}$ is clearly expressible as a gradient
\begin{equation}
\delta {\bf r}= 
\mbox{\boldmath $\nabla$} W ,
\end{equation}
where the function $W$ satisfies the equation
\begin{equation}
m \omega^2 W = \mbox{\boldmath $\nabla$} W  \cdot 
\left({\cal K}_0 {\bf r} \right)-
\left[ \mu- U_0({\bf r}) \right] 
\triangle W .\label{e_W}
\end{equation}
Expanding the solution (\ref{e_gensoln}) to first order, we obtain
\begin{equation}
\Phi\left({\bf r},t\right) \approx e^{-i \mu t/\hbar} 
\left[ \Phi_0\left({\bf r}\right) + 
\epsilon \left( u e^{-i \omega t}- v^{\ast} e^{i \omega t} \right) \right]
,
\label{e_pertsoln}
\end{equation}
where
\begin{equation}
u=\frac{f_+ + f_-}{2} , \, \, \, v=\frac{f_+ - f_-}{2}
\end{equation}
and 
\begin{equation}
f_{\pm}  =  C_{\pm}
\left[ 1-\frac{U_0\left({\bf r} \right)}{\mu} \right]^{\pm 1/2}  W ,
\end{equation}
with $C_-/C_+ = \hbar \omega /2 \mu  $. 
The expression (\ref{e_pertsoln}), with
the differential equation (\ref{e_W}) for $W$, is precisely the
result obtained by   \"Ohberg {\em et al.} \cite{Shlyapnikov}.
We have thus shown  that their solution describes the elementary phonon-like
excitations of the nonlinear hydrodynamical model.

The oscillatory modes for which the function $W({\bf r})$ is of third
or higher order in the components of ${\bf r}$ cannot be excited by 
linear deformations of the harmonic trap. That is, they cannot be produced
by applying a time-dependent harmonic potential of the form
(\ref{e_genpot})
to the ground state condensate $\Phi_0 \left({\bf r} \right)$.
However, the modes for which $W({\bf r})$ is linear or quadratic can be
produced in this way,
and for these modes it is possible to go beyond the perturbative limit,
to obtain a simple nonlinear matrix equation that fully describes
the dynamics of the condensate in the hydrodynamical regime. 

If the condensate is initially in the ground state
 $\Phi_0 \left({\bf r} \right)$, then 
the effect of a time-dependent
harmonic potential  (\ref{e_genpot}) 
 is simply to distort the condensate in a linear fashion
\begin{equation}
{\bf r}\left({\bf r}_0,t \right) = {\bf R}(t) + {\cal M}(t) {\bf r}_0 .
\end{equation}
Here ${\bf R}(t)$ is the centre of mass of the
condensate, and ${\cal M}(t)$ describes the time-dependent contraction
or shearing of the condensate along various directions.
The matrix $\partial {\bf r}/ \partial {\bf r}_0 $ is now 
independent of position
\begin{equation}
\frac{\partial {\bf r}}{\partial {\bf r}_0} = {\cal M}(t) =
 {\text{constant}} ({\bf r}_0) ,
\end{equation}
and equation (\ref{e_genevol}) can be separated into a part that is independent
of position, and a 
part that depends linearly on position, giving the following
two relations
\begin{eqnarray}
m \ddot{\bf R}& =& - {\cal K}(t) \left[{\bf R}-
 \bar{\bf r}(t)\right] \label{e_COM}  \\
m \ddot{\cal M} & = & - {\cal K}(t) {\cal M} + 
 \det \left( {\cal M} \right)^{-1}  \left( {\cal M}^{-1} \right)^T
{\cal K}_0  . \label{e_evM}
\end{eqnarray}
From equation (\ref{e_COM}) it is clear that for a stationary
trap (${\cal K}(t) ={\cal K}_0 $)
 the centre of mass motion
is separable along the three axial directions, and
 the frequencies of oscillation along these directions are
independent of amplitude, in agreement with the Kohn Theorem (see
 \cite{Dobson} and references therein). 

The velocity  can be written in terms of ${\bf r}$ as
\begin{equation}
{\bf v} =\dot{\bf R} + 
 \dot{\cal M} {\cal M}^{-1}\left( {\bf r}-{\bf R} \right) ,
\end{equation}
and hence the requirement that the velocity field have zero curl
is equivalent to the constraint that the matrix $\dot{\cal M} {\cal M}^{-1}$
be symmetric
\begin{equation}
\left[\dot{\cal M} {\cal M}^{-1}\right]^T = \dot{\cal M} {\cal M}^{-1} .
\label{e_constraint}
\end{equation}
 Using equation (\ref{e_evM}) it is easy to verify that
 provided condition (\ref{e_constraint})
is satisfied at the initial time it will be satisfied at all later
times.

Given this constraint  it is clear that
 for a stationary potential in the limit of small excitations, there must be
exactly six linearly independent modes for which the function $W({\bf r})$
is quadratic in the components of ${\bf r}$. For cylindrically
symmetric traps ($\omega_x = \omega_y \equiv \omega_{\perp}$) 
 they are (in the notation of reference
\cite{Shlyapnikov})  $\left\{ n=2,m=0, (\pm) \right\}$,
 $\left\{ n=0,m=\pm 2 \right\}$ and  $\left\{ n=1,m=\pm 1 \right\}$ 
\cite{footnote}. In this small amplitude limit
we can write 
\begin{equation}
{\cal M}(t) = 1 + \epsilon \left[ \delta {\cal M} e^{-i \omega t}+
\delta {\cal M} e^{i \omega t}\right],
\end{equation}
where $\delta {\cal M}$ is a constant matrix, independent of both position
and time. It completely characterises the modes and their linear
combinations in the small-amplitude limit. For the scaling modes
described by Castin and Dum \cite{CastinDum}
 (the modes $\left\{ n=2,m=0, (\pm)\right\}$, and
a symmetric superposition of the modes $\left\{ n=0,m=\pm 2 \right\}$) 
 $\delta {\cal M}$
is diagonal. For the mode $\left\{ n=0,m= +2 \right\}$, which we shall consider
next, it is

\begin{eqnarray}
\delta {\cal M} =  
  \left( 
     \begin{array}{ccc}
         1 &  i & 0 \\
         i & -1 & 0 \\
         0 &  0 & 0  
     \end{array} 
  \right) .
\end{eqnarray}
%

An exact solution for this mode can be found by setting 
${\cal K}(t) = {\cal K}_0$ in equation (\ref{e_evM}), and
is given by
\begin{equation}
{\cal M}(t) = \exp\left[\lambda {\cal Q}(t) \right] {\cal R}_{\lambda}(t) , \label{e_m2soln}
\end{equation}
where
\begin{equation}
{\cal Q}(t) = \left( \begin{array}{ccc}
\cos \left( \omega t \right) & \sin \left( \omega t \right) & 0 \\
\sin \left( \omega t \right) & -\cos \left( \omega t \right) & 0 \\
0 & 0 & 0  \end{array} \right), \label{e_cQ}
\end{equation}
and $\omega$ is the mode frequency. From expression (\ref{e_cQ}) 
for ${\cal Q}(t)$
it can be seen that the  transformation 
$\exp\left[\lambda {\cal Q}(t) \right]$  describes the distortion of the
condensate into an ellipse, and the rotation of that ellipse about
the $z$ axis. The parameter $\lambda$ is a measure of the amplitude of
excitation, and is precisely the quantity we labelled $ \epsilon$ in the small
amplitude limit.

The matrix ${\cal R}_{\lambda}(t)$ describes a slow rotation about the $z$
axis, together with a centrifugal 
stretching in the $x$-$y$ plane, and a corresponding
contraction along $z$
\begin{equation}
{\cal R}_{\lambda}(t) =  \left( \begin{array}{ccc}
\beta \cos \left( \omega_s t \right) & -\beta \sin \left( \omega_s t \right) & 0 \\
\beta \sin \left( \omega_s t \right) &\beta  \cos \left( \omega_s t \right) & 0 \\
0 & 0 & \beta^{-2/3}  \end{array} \right) .
\end{equation}
 This rotation is required  in order 
for ${\cal M}(t)$ to satisfy condition (\ref{e_constraint}).
 ${\cal R}_{\lambda}(t)$ depends on $\lambda$ through both the parameter 
$\beta$ and the slow frequency $\omega_s$. 

The aspect ratio $\alpha$ of the condensate is defined as
\begin{equation}
\alpha \equiv \frac{L_l - L_s}{L_l + L_s} \, ,
\end{equation}
where $L_l/L_s$ is the ratio of the lengths of the long and short axes
of the condensate in the $x$-$y$ plane. It is related to
the parameter $\lambda$ by
$
\alpha = \tanh \lambda .
$
The constant $\beta$ depends in turn on $\alpha$  through
$
\beta =\lbrack (1 + \alpha^4)/(1 - \alpha^4) \rbrack^{3/10} .
$
The slow frequency $\omega_s$ is given by
$
\omega_s = \lbrack \sqrt{2}\alpha^2/\sqrt{1+\alpha^4}\rbrack
 \omega_{\perp} ,
$
and the mode frequency $\omega$ scales with the aspect ratio as
\begin{equation}
\omega = \sqrt{2} \frac{ 1+ \alpha^2 }
{\sqrt{1+\alpha^4}}\omega_{\perp} .
\end{equation}
$\omega$ thus depends on the amplitude of
excitation.
For very low amplitudes, 
$\lambda \rightarrow 0$,
$\alpha \rightarrow 0$, and the mode frequency tends to
$\omega \rightarrow \sqrt{2}\omega_{\perp} $.
At very high amplitudes, 
$\lambda \rightarrow \infty$, $\alpha \rightarrow 1$
and the mode frequency approaches the limit
$\omega \rightarrow 2 \omega_{\perp}$, which coincides with the
frequency of oscillation of a non-interacting gas.

This mode has been observed experimentally by 
 Jin {\em et al.} \cite{Cornell_1,Cornell_2}, who measured
its frequency as a function of response amplitude
(aspect ratio) after opening the
trap and allowing the condensate to expand. Due to the presence of
atomic interactions, the aspect ratio will change during the
expansion. In order to compare our results with the
experimental data it is therefore necessary to model
the evolution of the condensate after the trap has been
opened. We assume that the opening of the trap occurs
instantaneously. Taking the solution 
(\ref{e_m2soln}) for the trapped condensate as an initial condition, 
we insert it into (\ref{e_evM}), setting ${\cal K}(t)=0$, and integrate
numerically.
 The aspect ratio
of the condensate is calculated from the elements of ${\cal M}$ after
the appropriate expansion time.
\begin{figure}
\begin{center}
\leavevmode
\epsfxsize=0.45\textwidth
\epsffile{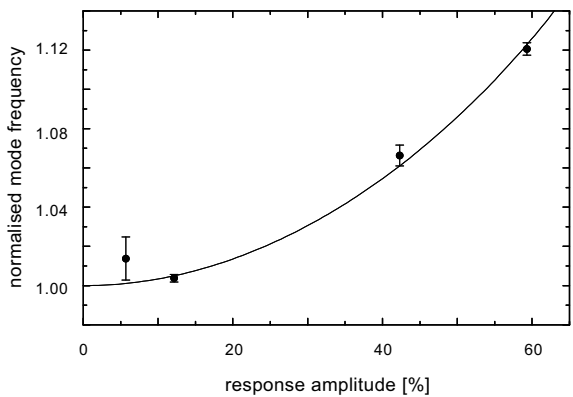}
\end{center}
\caption
{
The normalised mode frequency $\omega/\omega(0)$  is plotted
against response amplitude (aspect ratio $\alpha$) after expansion. 
Our theoretical predictions (curve) are compared with the experimental data of
Jin {\em et al.} \protect\cite{Cornell_2}.
\label{f_figure1}
}
\end{figure}
Shown in Figure 1 is the experimental
data obtained by Jin {\em et al.}  \cite{Cornell_2}
for the mode frequency as a function of aspect ratio.
These points were obtained with
a radial trap frequency $\omega_{\perp}/2 \pi$ 
of 132 Hz and an expansion time of 7 ms \cite{Cornell_1}, and
the frequencies were normalised to the small amplitude value $\omega(0)$
\cite{PrivateCommunication}. The ratio
of the axial to radial trap frequencies was 
$\omega_z / \omega_{\perp} = \sqrt{8}$. The
number of condensed atoms was approximately 4500, and  temperature
was $T=(0.50 \pm 0.08) T_c$, where $T_c$ is the critical temperature.
The theoretical curve, calculated for the same parameters,
shows good agreement with the experimental data.

Note that in our treatment we totally neglected 
the the presence of the non-condensed 
particles at a finite temperature $T$. Within the mean-field 
framework the force caused by the non-condensed particles can be shown
to be $(T/T_{c})^{3/2} (\mu/T_{c})^{3/2}$ times smaller than the one 
related to the condensate, which makes this assumption valid 
up to a fraction of the transition temperature.  


In conclusion, we have considered the collective excitations of
a Bose-Einstein condensate in the hydrodynamical regime. We have shown
that the excitations induced by arbitrary time-dependent linear deformations of the 
trapping potential form a broad class of solutions of the
hydrodynamical equations, which remains closed under evolution. The
dynamics of the condensate for this class of excitations can be described
by a  nonlinear equation involving a $3 \times 3$ matrix. We have
obtained an exact solution of this equation for the  $\left\{ n=0,m=+2 \right\}$
mode, which shows good agreement with experimental results.

We would also like to note that described in the literature 
3-fold class of nonlinear solutions \cite{CastinDum} covers  
two $\left\{ n=2,m= 0 \right\}$ modes and any linear combination 
of $\left\{ n=0,m=\pm 2 \right\}$ modes with (generally complex) 
{\it equal by the absolute value}
coefficients: the observed in the experiment \cite{Cornell_2} 
$\left\{ n=0,m=+2 \right\}$ mode does not belong to this class.  
The 6-fold closed class of nonlinear solutions obtained 
in our work extends the treatment to two other modes 
($\left\{ n=0,m=\pm 2 \right\}$ respectively) as well as 
to an {\it arbitrary} superposition of $\left\{ n=0,m=\pm 2 \right\}$
modes, including $\left\{ n=0,m=+2 \right\}$.

In the other words, the existing treatment \cite{CastinDum}
can describe any excitation obtained by a modulation of the
matrix elements of the spring-constant tensor ${\cal K}_{ij}(t)$
(see (\ref{e_genpot})) provided it always remains {\it diagonal}.
This model can not be applied to describe the rotation-like modulations
such as the ``$m=2$'' JILA experiment \cite{Cornell_2}. 
Instead, 
our paper presents an extention for this model which can be applied to
{\it any} kind of the time dependence of the spring-constant tensor,
including the one used in the above experiment.

M.O.\ was supported by the National Science Foundation
grant for light force dynamics \#PHY-93-12572. 
 P.S.\ was supported
by the University of Auckland. She is grateful to Professor R.\ Glauber
for making possible her visit to Harvard.
This work was also partially supported by
the NSF through
a grant for the Institute for Theoretical Atomic and Molecular
Physics at Harvard University and the Smithsonian Astrophysical 
Observatory. The authors are grateful to Professor E.\ Cornell
for helpful comments.
%
%

%


\begin{thebibliography}{99}
\bibitem{footnote}
Note that in the limit of a spherically symmetric trap, the `fast mode'
 $\left\{ n=2,m=0, (+) \right\}$ reduces to the $l=0$
mode with frequency $\sqrt{5}\omega_{\text{trap}}$, and the remaining
five modes form an $l=2$ quintuplet with frequency
 $\sqrt{2}\omega_{\text{trap}}$.
\end{thebibliography}
\end{document}